\documentclass[structabstract]{aa}
\usepackage{graphicx}
\usepackage{txfonts}
\usepackage{natbib}
\usepackage{color}
\usepackage[font=small,format=plain,labelfont=bf,up,textfont=it,up]{caption}

\begin{document}
\title{The magnetic field of the evolved star W43A}

\author{N. Amiri
      \inst{1,3}
      \and
       W. Vlemmings
       \inst{2}
        \and
        H.J. van Langevelde
        \inst{3,1}
        }
  \institute{Sterrewacht Leiden, Leiden University, Niels Bohrweg 2, 2333 CA Leiden, The Netherlands
            \and
            Argelander Institute for Astronomy, University of Bonn, Auf dem H\"{u}gel 71, 53121 Bonn, Germany
            \and
            Joint Institute for VLBI in Europe (JIVE), Postbus 2, 7990 AA Dwingeloo, The Netherlands
            }
     \date{Received 28 August  2009 / Accepted 13 October 2009}
\abstract 
{ The majority of the observed planetary nebulae exhibit elliptical or bipolar structures. Recent observations have shown that asymmetries already start during the last stages of the AGB phase. Theoretical modeling has indicated that magnetically collimated jets may be responsible for the formation of the non-spherical planetary nebulae. Direct measurement of the magnetic field of evolved stars is possible using polarization observations of different maser species occurring in the circumstellar envelopes around these stars.}{The aim of this project is to measure the Zeeman splitting caused by the magnetic field in the OH and H$_2$O maser regions occurring in the circumstellar envelope and bipolar outflow of the evolved star W43A. We compare the magnetic field obtained in the OH maser region with the one measured in the H$_2$O maser jet.}{We used the UK Multi-Element Radio Linked Interferometer Network (MERLIN) to observe the polarization of the OH masers in the circumstellar envelope of W43A. Likewise, we used the Green Bank Telescope (GBT) observations to measure the magnetic field strength obtained
previously in the H$_2$O maser jet.}
{We report a measured magnetic field of approximately 100 $\mu G$ in the OH maser region of the circumstellar envelope around W43A. The GBT observations reveal a magnetic field strength $B_{\rm ||}$ of $\sim$30 mG changing sign across the H$_2$O masers at the tip of the red-shifted lobe of the bipolar outflow. We also find that the OH maser shell shows no sign of non-spherical expansion and that it probably has an expansion velocity that is typical for the shells of regular OH/IR stars.} {The GBT observations confirm that the magnetic field collimates the H$_{2}$O maser jet, while the OH maser observations show that a strong large scale magnetic field is present in the envelope surrounding the W43A central star. The magnetic field in the OH maser envelope is consistent with the one extrapolated from the H$_2$O measurements, confirming that magnetic fields play an important role in the entire circumstellar environment of W43A.}

\maketitle
\section{Introduction}\label{introduction}
At the end of their evolution, low mass stars undergo a period of high mass loss ($\dot{M} \propto 10^{-4} -  10^{-7}$  M$_{\odot}$/yr) that is important for enriching the interstellar medium with processed molecules. During this stage the star climbs up the asymptotic giant branch (AGB)  in the Hertzsprung-Russell (H-R) diagram. AGB stars are generally observed to be spherically symmetric \citep{habingreview}. However, planetary nebulae (PNe), supposedly formed out of the ejected outer envelopes of AGB stars, often show large departures from spherical symmetry. The origin and development of these asymmetries is not clearly understood. Observations of collimated jets and outflows of material in a number of PNe have been reported \citep[e.g][]{sahai1998, miranda2001, alcolea2001}.  \cite{sahai1998} propose that the precession of these jets is responsible for the observed asymmetries. These jets are likely formed when the star leaves the AGB and undergoes a transition to become a PN \citep[e.g.][]{sahai1998,imainature,miranda2001}. 

Magnetic fields can play an important role in shaping the circumstellar envelope (CSE) of evolved stars and can produce asymmetries during the transition from a spherical symmetric star into a non-spherical PN. They are also possible agents for collimating the jets around these sources \citep{garcia2005}. It is not clear how stars can maintain a significant magnetic field throughout the giant phases, as the drag of a large scale magnetic field would brake any stellar dynamo if no additional source of angular momentum is present \citep[e.g.][]{soker1998}. However, theoretical models have shown that AGB stars can generate the magnetic field through a dynamo interaction between the fast rotating core and the slow rotating envelope \citep{blackmannature}. Alternatively, the presence of a heavy planet or a binary companion as the additional source of angular momentum can maintain the magnetic field \citep[e.g.][]{frank2004}. 

Polarization observations of different maser species in the CSE of these stars provide a unique tool for understanding the role of the magnetic field in the process of jet collimation. Strong magnetic fields have been observed throughout the entire CSE of these stars for different molecular species (e.g. \citealt{etoka2004, wouternature,herpin2006} using OH, H$_{2}$O and SiO masers, respectively).

\citet{likkel1992} introduced a separate class of post-AGB sources where H$_{2}$O maser velocity range exceeds that of OH maser in the CSE. High resolution H$_{2}$O maser observations trace highly collimated jets in the inner envelopes of these objects. These objects, the so-called water fountain sources, are thought to be in the transition stage to PNe. An archetype of this class is W43A, located at a distance of 2.6 kpc from the sun \citep{diamond1985}, for which the H$_2$O masers have been shown to occur at the tips of a strongly collimated and precessing bipolar jet \citep{imainature}. Polarization observations of the H$_2$O masers of W43A reveal a strong magnetic field apparently collimating the jet \citep{wouternature}.

Here we describe observations of the Zeeman splitting of OH masers in the CSE and H$_{2}$O masers in the jet of W43A. Our aim is to investigate the magnetic field strength and morphology in the low density region of the CSE of this object, which is in transition to become a PN and confirm the magnetic field in the H$_2$O maser jet. The observations are described in \S \ref{observations} and the results of the  observed spectra and the measured Zeeman splitting are presented in \S \ref{results}. The results are discussed in \S \ref{discussion}, where we investigate the observed Zeeman splitting and the significance of other non-Zeeman effects. This is followed by conclusions in \S \ref{conclusion}.

\section{OBSERVATIONS AND DATA ANALYSIS}\label{observations}
\subsection{MERLIN Observations}
We used the UK Multi-Element Radio Linked Interferometer Network (MERLIN) on 4 June 2007 to observe the 1612 MHz OH masers of the evolved star W43A. We included the Lovel telescope to achieve a higher sensitivity in this experiment. The longest baseline (217 km) resulted in a beam size of 0.3 $\times$ 0.2 arcsec. The observations were performed in full polarization spectral line mode with the maximum possible 256 spectral channels and a bandwidth of 0.25 MHz, covering a velocity width of 44 km/s, this gives a channel width of 0.2 km/s. 

The observations of W43A were interleaved with observations of the phase reference source, 1904+013, in wide-band mode in order to obtain an optimal signal-to-noise-ratio. 3C286 was observed as primary flux calibrator and polarization angle calibrator. 3C84 was observed both in narrow band and wide band modes in order to apply band pass calibration and a phase offset correction. The full track of observations of W43A was 10 hours. 

We performed the initial processing of the raw MERLIN data and conversion to FITS using the local d-programs at Jodrell Bank. The flux density of the amplitude calibrator, 3C84, was determined using the flux density of the primary flux calibrator, 3C286; we obtained a flux density of 18.94 Jy for 3C84. The rest of the calibration, editing and reduction of the data were performed in the Astronomical Image Processing Software Package (AIPS). Since the wide and narrow band data cover different frequency
       ranges and use different electronics, there may be a phase offset between the wide-band and narrow-band data; this offset was measured using the difference in phase solutions for 3C84 and was applied to correct the phase solutions from the phase reference source. The phase reference source, 1904+013, was used to obtain phase and amplitude solutions, which were applied to the target data set. The polarization calibration for leakage was done using 3C84, and the R and L phase offset corrections were performed on 3C286. Image cubes were created for stokes I, Q, U and V. The resulting noise in the emission free channels was 6.5 mJy/Beam.  For the brightest feature we are 
limited to a dynamic range of $\sim$650. A linearly polarized data cube was made using the stokes Q and U.
\subsection{GBT Observations}
The observations of the H$_{2}$O masers in the bipolar outflow of W43A
were carried out at the NRAO\footnote{The National Radio Astronomy
Observatory (NRAO) is a facility of the National Science Foundation
operated under cooperative agreement by Associated Universities, Inc.}
GBT at Oct 14 2006 as part of a project aimed at detecting additional
water fountain sources. At 22.2~GHz the full-width at half maximum
(FWHM) beamwidth of the GBT is $\sim 33''$, while the H$_2$O maser
emission is located within $\sim 1''$. The GBT spectrometer was used
with a bandwidth of 200~MHz and 16,384 spectral channels, providing a
channel spacing of $0.164$ km/s and a total velocity coverage of
$2700$ km/s, centered on $V_{\rm LSR}=34$ km/s. Furthermore, the data
were taken with the dual-polarization receiver of the lower $K$-band
using the total power nod observing mode. The two beams have a fixed
separation of $178.8''$ in the azimuth direction and a cycle time of 2
minutes was sufficient to correct for atmospheric variations. As a
result, one beam of the telescope was always pointing at the source
while the other beam was used for baseline correction. As W43A was
only observed as a test for the spectral line setup, the total on
source observing time was just 6 minutes. Pointing and focus observations were
done on J0958+655. This source, and 3C286, were also used as flux
calibrator, providing flux calibration accurate to $\sim10\%$. For
data reduction we used the GBT
IDL\footnote{http://gbtidl.sourceforge.net} software package. The
total intensity spectrum of the W43A H$_2$O masers from these
observations is shown in Fig. \ref{fig:OHandGBT}.

\subsection{Determining Zeeman Splitting}
The Zeeman effect causes a velocity shift between the left circular polarization (LCP) and the right circular polarization (RCP) spectra.
For  OH masers the separation is often larger than the line width, but for the  H$_{2}$O molecule the splitting is small. The cross-correlation introduced by \citet{modjaz2005} is an effective technique for measuring the Zeeman splitting without forming the stokes V spectrum. In this method, the RCP and LCP spectra are cross-correlated to determine the velocity offset. This method can even work for complex spectra, assuming that the velocity offset is the same over the spectrum; which means the magnetic field strength and direction is constant in the masing region. The sensitivity of this method is comparable to the S-curve method, where the stokes V spectrum is directly used for measuring the magnetic field \citep[e.g.][]{vlemmings2001,fiebig1989}, but has the advantage of being less susceptible to calibration errors in the RCP and LCP absolute flux level determination.

\section{Results}\label{results}

\begin{figure}
\centering
  \resizebox{\hsize}{!}{\includegraphics[scale=0.5]{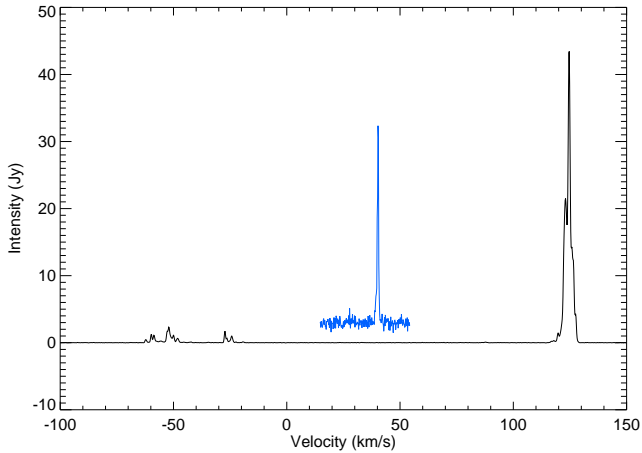}}
  \caption{The OH and H$_{2}$O maser spectra of W43A. The blue curve is the MERLIN 1612 MHz OH maser emission spectrum which is shifted upward  by 2 Jy for the purpose of illustration. The black curve is the H$_{2}$O maser emission at 22 GHz obtained from GBT observations.}
  \label{fig:OHandGBT}
\end{figure}

Fig. \ref{fig:OHandGBT} shows the spectrum of the OH and H$_{2}$O  maser regions of W43A obtained from MERLIN and GBT observations, respectively. For W43A, the velocity range for the OH (27 to 43 km/s) is much less than for H$_{2}$O (-53 to 126 km/s). The figure illustrates the water fountain nature of this source where H$_{2}$O masers occur outside the OH maser region in a much larger velocity range. The W43A OH masers have previously been observed by \citet{bowers1978} with a bandwidth of $\sim$ 400 km/s, who reported the OH emission to be confined to a similar velocity range.

\begin{figure}
  \resizebox{\hsize}{!}{\includegraphics[scale=0.3]{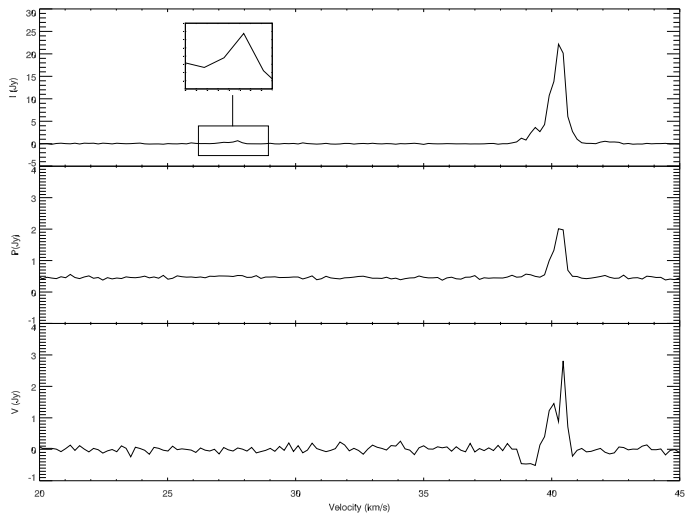}}
  \caption{The 1612 MHz spectra for total intensity (I), linear polarization (P) and circular polarization (V) data.}
  \label{fig:I spectra}
\end{figure}

 The OH velocity profiles of the integrated flux of each channel in the I (total intensity), V (circular polarization) and P (linear polarization intensity) data are shown in Figure \ref{fig:I spectra}. The brightest peak is red-shifted and the blue-shifted peak has a much lower brightness; only $ \sim \ 3 \% $ of the red-shifted brightness. There is little emission detected between the two peaks. Most of the emission in the total intensity profile was also detected in the linear and circular polarization spectra. For the circular and linear polarization profiles, the emission is dominated by the red-shifted peak. The peaks in the polarization intensity and circular polarization spectra are 10$\%$ linearly and 12$\%$ circularly polarized. 

Using the AIPS task SAD, OH maser features with peaks higher than three times the rms noise in the emission free channel were identified and fitted with a Gaussian in the total intensity image cube; the results of which are shown in Table \ref{tab:results}. The maser line width for each feature ($\Delta {\nu}_{l}$), was obtained by fitting a Gaussian distribution to the I spectra. The Zeeman splitting ( $\Delta {\nu}_{z}$) is the velocity shift between the right and left circular polarizations for each maser feature obtained from the cross correlation method. B is the value of magnetic field measured by converting the velocity splitting into the magnetic field strength using the Zeeman splitting coefficient of the OH maser line taken from the literature \citep[236 km s$^{-1}$ G$^{-1}$;][]{davies1974}. The errors in field determination and Zeeman splitting depend on the rms in the channels with bright emission. As the noise in channels with bright emission increases by a factor of  $\sim$5, we have conservatively determined the errors by the increased rms in the channels with strong maser signal. The robustness of the errors determined from this against measurement uncertainties has been discussed in depth by \citet{modjaz2005}. The fractional linear (m$_{l}$) and circular (m$_{c}$) polarizations were obtained from the polarization intensity and circular polarization spectra and $\chi$ denotes the polarization angle.
\begin{table*}
\begin{center}
\begin{tabular}{|c |c |c |c |c |c|c|c|c|c|c|}
\hline
Feature &RA&Dec& Peak Flux  & $V_{rad}$ & $\Delta \nu _ {l}$ &$\Delta \nu _ {z}$&B  & m$_{l}$ &m$_{c}$&$\chi$\\ 
&18 47 &-1 05 &(Jy beam$^{-1}$)&(km s$^{-1}$)&(km s$^{-1}$)&(m s$^{-1}$)&($\mu$G)&$\%$&$\%$&deg\\
\hline\hline
1 &41.15933&11.4200&$1.08 \pm 0.01$ & 41.0& 0.59 &$11.9\pm 0.2$& 50$\pm$4& - &$4.6 \pm 0.9$& -\\ 
2 &41.15684&11.4500&$2.18 \pm 0.01$ & 40.8& 0.59 &$13.0\pm 0.2$& 55$\pm$4 &$1.1 \pm 0.3$   &$11.5 \pm 0.4$& -11$\pm$1\\ 
3 &41.15676&11.4694&$5.20 \pm 0.01$ & 40.6& 0.59 &$13.9\pm 0.3$& 59$\pm$6 &$0.8 \pm 0.2$&$5.0 \pm 0.2$& -17$\pm$7\\ 
4 &41.15666&11.4600&$21.04 \pm 0.03$& 40.2& 0.58 &$13.9\pm 0.2$& 59$\pm$4 &$3.13 \pm 0.05$&$3.9 \pm 0.1$& -4.3$\pm$0.4\\ 
5 &41.15952&11.5428&$1.97 \pm 0.01$ & 39.5& 0.62 &$20.1\pm 0.5$& 85$\pm$11 &-&$2.6 \pm 0.5$& -\\ 
6 &41.15938&11.5790&$3.33 \pm 0.01$ & 39.3& 1.54 &$39.1\pm 1.1$& 166$\pm$24& $1.5 \pm 0.3$ &$14.7 \pm 0.3$& -0.44$\pm$5.7\\ 
7 &41.17281&11.3021&$0.20 \pm 0.01$ & 27.9& 0.43 &-& - & - &-&-\\ 
8 &41.17217&11.3431&$0.60 \pm 0.01$ & 27.7& 0.60  &- &-& $6.6 \pm 1.32$ &-&-44$\pm$7\\ 
\hline
\end{tabular}
\caption{OH maser Results}
\label{tab:results}
\end{center}
\end{table*}

\begin{figure}
  \resizebox{\hsize}{!}{\includegraphics[scale=0.8]{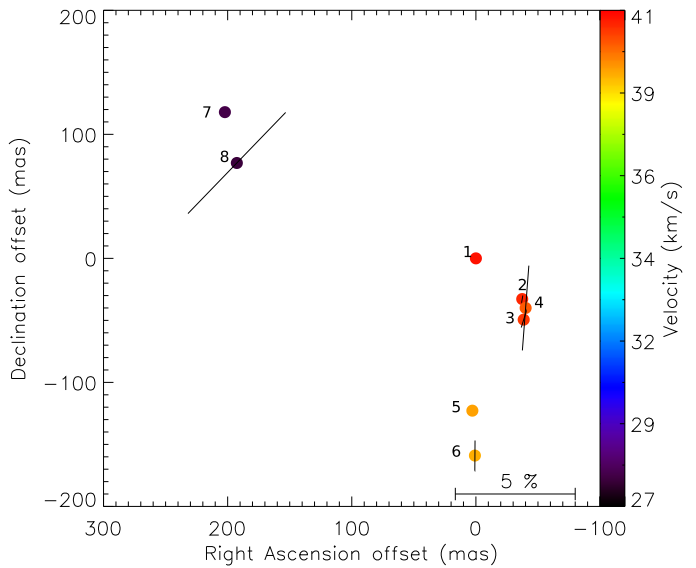}}
  \caption{The OH maser region of W43A. The offset positions are with respect to the maser reference feature. The maser spots are col-our-coded according to radial velocity. The vectors show the polarization angles, scaled logarithmically according to the linear polarization fraction m$_{l}$ (Table \ref {tab:results}).}
  \label{fig:maser spots}
\end{figure}

Fig. \ref{fig:maser spots} shows the OH maser spots (red-shifted and blue-shifted). The OH maser features detected for W43A are color coded according to their radial velocity (Table \ref{tab:results}). The vectors indicate the polarization angle scaled logarithmically according to fractional linear polarization. The weighted average of the red-shifted vectors is $\sim$ -6.4$^{\circ}$ while the polarization angle for the blue-shifted feature is -44 $^{\circ}$.

For illustration, following Fig. 2 of \citet{imainature}, Fig. \ref{fig:maser distribution} shows the spatial distribution of the OH maser features of W43A together
with H$_{2}$O maser positions \citep{wouternature}. Since no accurate astrometry was available for H$_{2}$O masers, the maps were aligned on the respective centers of the maser distribution.

\begin{figure}
 \resizebox{\hsize}{!}{\includegraphics{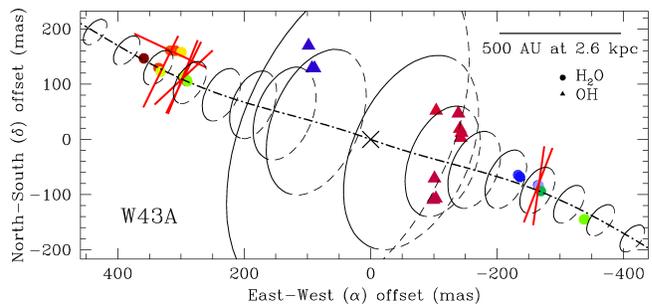}}
 \caption{Spatial distribution of OH and H$_{2}$O maser features. The offset positions are with respect to the maser reference feature. H$_{2}$O maser features are indicated by filled circles. OH maser features are shown as triangles. Red and blue colors show the
red-shifted and blue-shifted features. }
 \label{fig:maser distribution}
\end{figure}

\section{Discussion}\label{discussion}
\subsection{OH maser polarization}
The linear and circular polarization that we measure (as illustrated in Table 1), is consistent with  marginal detections reported from earlier observations \citep{wilson1972}. They found $4 \pm 2 \%$ linear polarization for the masers at $\sim$40.5 km/s, with a polarization angle of 10$\pm$20 degrees. This is consistent with our finding of 0.8-3.2$\%$ linear polarization and an average polarization angle of $\sim$-7 degrees for the masers around that velocity. Additionally, they measured a peak circular polarization of $30 \pm 10   \% $ at $\sim$ 39 km/s, decreasing to 10$\pm$5$\%$ towards 40 km/s, which is consistent with our observations.

The Zeeman splitting for the OH is larger than that for H$_{2}$O. The split energy is determined by the following equation:
\begin{equation}
 \Delta W \ = \ -\mu _{0} \  g \ m_{F} \  B.
\end{equation}
For the OH molecule $\mu _{0}$ is the Bohr magneton ($\mu _{B} = \frac {e \hbar} {2 m_{e} c}$), and for the H$_{2}$O molecule $\mu _{0}$ is the Nuclear magneton ($\mu _{N} = \frac {e \hbar} {2 m_{n} c}$). As the Bohr magneton is almost 3 orders of magnitudes larger than the Nuclear magneton, the OH Zeeman splitting is correspondingly larger. However, the observed Zeeman splitting in the OH maser region of W43A is considerably still less than the maser line width by a factor of 50 (Table \ref{tab:results}). We obtained an average magnetic field of 100 $\mu$ G in the OH maser region of W43A, using the cross correlation between the LCP and RCP spectra. The measured magnetic field is an order of magnitude lower than those often found in OH  maser regions of evolved stars \citep[e.g.][]{etoka2004,bains2003}. However, \cite{zell1991} reported magnetic field on micro-gauss level in the envelopes of a number of OH/IR stars. They also argue that there is convincing evidence that  the smoothness of the line profile is consistent with models in which there are a few thousand individual emitting elements, with 5 or 10 individual elements within the spectral resolution. This implies that the spectral blending could decrease the observed polarization by as much as a factor of 2-3. Likewise, a comparison between high- and low spatial resolution circular polarization observations of other maser species also indicates that the blending of maser features typically decreases the magnetic field measured at low angular resolution by a factor of 2 \citep[e.g.][]{sarma2001}. Interferometric observations with higher spatial and spectral resolution are required to explore this effect further. Unfortunately, many of the masers may be resolved out due to their extended structure. Alternatively, OH polarization could originate from non-Zeeman effects. Although our measured magnetic field strength is similar to what is reported reported in OH/IR envelopes \citep{zell1991}, we investigate below to what level these effects could be contributing to our polarization measurements.

\subsubsection{Non-Zeeman effects}

A possible source of intensity dependent circular polarization occurs because of  the rotation of the axis of symmetry for the molecular quantum states \citep{nedoluha1994}. This stems from  the competition between the magnetic field (B) and the rate of stimulated emission (R). While g$\Omega \geq \ R$, the magnetic field is the quantization axis. However, as the maser saturates and the rate of stimulated emission becomes larger, the molecule interacts more strongly with the stimulated emission and the axis of the symmetry of the molecule changes toward the direction of the maser beam. For the 1612 MHz OH maser the Zeeman splitting is 1.308 MHz G$^{-1}$ \citep{davies1974}. For a magnetic field of B $\geq \ 10 \ \mu$G, the Zeeman frequency shift becomes  g$\Omega \geq \ 13.08$ s$^{-1}$. The rate of stimulated emission is:

\begin{equation}
 R \simeq \ A \  K  \  T_{b}  \   \Delta\Omega  \  /   \  4  \   \pi  \   h  \   \nu.     
\end{equation}

Here A is the Einstein coefficient for maser transition which is $1.3 \times 10^{-11}  \  s^{-1}$ for the 1612 MHz OH maser emission \citep{destombes1977}. K and h are the Boltzmann and Planck constants respectively. $T_{b}$ denotes the brightness temperature and for the 1612 MHz OH maser the maximum value is $10^{11}$ K \citep{reid1981}. $\Delta \Omega$ is the maser beaming angle. For a typical angle of $10^{-2}$ sr, the maser stimulated emission rate becomes 0.013 $s^{-1}$. Therefore, even for the largest value of R, the Zeeman frequency shift (g$\Omega$) is higher than the stimulated emission rate. The imposed change of the symmetry axis due to stimulated emission can thus not explain the circular polarization observed in the OH  maser region of W43A.

Alternatively, the propagation of a strong linear polarization can create circular polarization if the magnetic field orientation changes along the maser propagation direction \citep{wiebe1998}. In the unsaturated regime with a fractional linear polarization up to 50$\%$, the generated circular polarization is $\frac{{m_{l}}^{2}}{4}$ when the magnetic field rotates 1 rad along the maser path. In this case, the linear polarization fraction observed in the OH  maser region of W43A (m$_l$ $\sim$ 10 $\%$), implies a generated circular polarization (m$_c$ $\leq$ 0.25 $\%$), which is much less than the measured circular polarization. Thus it is highly unlikely that the observed circular polarization is created in this way in the OH shell of W43A.

Thirdly, \cite{fish2006} discuss an observational effect which may generate velocity shift between RCP and LCP in the presence of a large linear polarization fraction, which could be falsely attributed to Zeeman splitting. They consider an extreme case where the emission is right elliptically polarized. The linearly polarized flux will only appear in the LCP receiver and the RCP receiver picks up all the emission including linear and circular polarizations. If the magnetic field orientation changes along the amplification path, the linear polarization component may be shifted in velocity with respect to the circular polarized component. This offset will manifest itself as a velocity shift between the RCP and LCP spectra. However, in the case of the OH emission region of W43A the RCP and LCP are at the same intensity level which means that both contain linear and circular polarizations. Therefore, this effect is unlikely to be at work.

Finally, to investigate the effect of instrumental polarization on the measured magnetic field, we imaged the unpolarized source, 3C84, in all polarization states and obtained the fractional linear and circular polarization which may account as leakage. 3C84 is regularly used as  VLA/VLBA/MERLIN polarization calibrator, and is known to be unpolarized.  Our results indicate a limit of 3$\%$ linear and 1$\%$ circular instrumental polarization. The relative low level of instrumental polarization could therefore not generate the observed circular polarization of 12$\%$ in the OH maser region of W43A. However, we can not formally rule out instrumental polarization as the cause of the linear polarization observed for any of the maser features other than feature 8.

\subsection{H$_{2}$O maser polarization}
\begin{figure}
\centering
  \resizebox{\hsize}{!}{\includegraphics[scale=0.5]{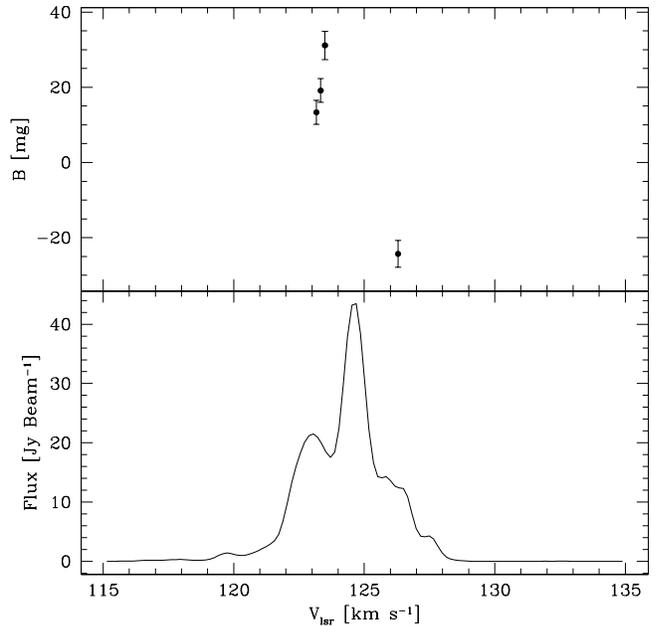}}
  \caption{Bottom panel: The total intensity spectrum of the red-shifted H$_2$O  masers of W43A obtained from the GBT observations. The top figure shows the magnetic field strength measured for the red-shifted part of the spectrum. }
  \label{fig:GBT}
\end{figure}

The magnetic field on the red-shifted H$_2$O masers of W43A was
detected using the cross-correlation method that was also used to
measure the magnetic field in the OH  maser region. To distinguish
between separate spectral features, we specifically used the 'running'
average method described in \citet{woutermethanol} over 3 km/s intervals. We
detect a magnetic field strength $B_{\rm ||}$ changing sign from
$31\pm8$~mG to $-24\pm7$~mG across the red-shifted maser region (Fig. \ref{fig:GBT}). This
is approximately a factor of two lower than the magnetic field
measured on blue-shifted masers using the VLBA \citep{wouternature}. However, blending of the maser features will decrease the
magnetic field measured with low angular resolution and small
differences of the pre-shock density, pre-shock magnetic field, or the shock velocity at blue and red-shifted tips of the jet will also
affect magnetic field strength in the shock compressed H$_2$O  maser region \citep{elitzur1989}. The measurements are thus in good
agreement with the previously published results. Additionally, the sign
reversal seen across the maser provides additional support for the
proposed jet collimation by a toroidal magnetic field as we would
expect the magnetic field to change sign on either side of the jet.

\subsection{The role of the magnetic field}
Taking into account all the possible effects which may contribute to the velocity offset between RCP and LCP, we found no significant effect other than Zeeman splitting that could explain the observed circular polarization of the OH maser of W43A. We conclude that the observed circular polarization has a Zeeman origin and implies the OH  maser region contains a large scale magnetic field of B $\sim$ 100 $\mu$G.

\citet{wouternature} observed the H$_{2}$O  maser region of W43A at 22 GHz and measured a magnetic field of 85$\pm$33 mG in the blue-shifted region. Our GBT observations reveal magnetic field strength $B_{\rm ||}$ changing sign from $31\pm8$~mG to $-24\pm7$~mG across the red-shifted  H$_{2}$O  masers. The magnetic field measured from the GBT observations confirms previous results by \citet{wouternature} that the magnetic field has a role in collimating the jet of W43A. H$_{2}$O masers occur in gas with a hydrogen number density of n $\approx$ $10^{8}-10^{10}$ cm $^{-3}$  and OH is excited in gas with a hydrogen number density of $10^{4}-10^{6}$ cm$^{-3}$ \citep{elitzur_1992}. Two different scenarios for H$_{2}$O masers in W43A exist. One possibility is that the masers occur at the tips of the jet, when the jet has swept up enough material previously ejected from the stellar atmosphere. The jet occurs at the distance of 1000 AU from the central star traced by H$_{2}$O maser observations  and the typical hydrogen density at the distance of 1000 AU is $10^{5}$ cm$^{-3}$ \citep{wouternature}. Therefore, the density must increase by three orders of magnitude at the tips of the jet so that the conditions become appropriate for H$_{2}$O maser excitation. Since the magnetic field strength depends on the density of the material (B $\propto n^{k}$, with magnetic field measurements in star forming regions implying k$\sim$0.5), the extrapolation of magnetic field outside the jet implies the value of B $\approx$ 0.9-2.6 mG in the OH  maser region \citep{wouternature}. Alternatively, the masers can occur in a shock between the collimated jet and the dense material outside the CSE, similar to the H$_{2}$O masers occurring in star forming regions \citep{elitzur1989}. Then, the pre-shock magnetic field extrapolated from the H$_{2}$O observations is 70 $\mu$G, comparable to what we now find for the OH maser region and implies a pre-shock density of  3$\times 10^{6}$ cm$^{-3}$  \citep{wouternature}. Previously, it was thought that the H$_{2}$O masers are likely to occur in the compressed material at the tips of the jet because the pre-shock density is somewhat higher than the expected value at 1000 AU from the star in the circumstellar envelope \citep{wouterapj}. However, our observations show that it is more likely that the H$_{2}$O masers occur in a shock since the measured magnetic field of 100 $\mu$G is more consistent with the estimated magnetic field from the shock model (70 $\mu$G).

In the H$_{2}$O  maser region of W43A, \citet{wouternature} concluded that the linear polarization is perpendicular to the magnetic field and aligned to the jet. However, at the low frequency of OH , Faraday rotation makes the derivation of the magnetic field configuration impossible. For a source at 2.6 kpc and a typical value of the interstellar magnetic field of 1 $\mu$G and a density of n$_{e}=0.03$ cm$^{-3}$, the Faraday rotation is $\sim$ 125 $^{\circ}$. Additionally, the internal Faraday rotations introduce large scattering of polarization angle \citep{fish2006}.
Therefore, it is not possible to determine the absolute geometry of the magnetic field in the OH maser region of W43A.

\subsection{OH maser shell expansion of W43A}
Our observations reveal that the OH and H$_{2}$O masers in the CSE of W43A occur in two emission clusters with opposite velocity separations. The locations of the blue and red-shifted OH maser components are reversed compared with the  H$_{2}$O emission features (Fig. \ref{fig:maser distribution}). 

From our MERLIN observations, we obtained an angular separation of 0.28$\pm$0.02 arcsec between the red and blue-shifted OH features. The observed velocity difference is $\sim$ 13 km/s between the two emission complexes. The fact that the red and blue-shifted features are not coincident on the sky is not fully compatible with a spherically symmetric expanding shell. Similar observations of the OH maser shell of W43A were performed previously. The 1612 MHz MERLIN observations by \cite{diamond1988} on March 1981, showed an angular separation of  0.21$\pm$0.03 arc sec with a velocity splitting of $\sim$ 16 km/s between the two emission clusters. Assuming the two sets of observations are tracing the same sites of OH emission, the measured expansion is 0.07 $\pm$0.03 arcsec in 26.5 years, which is equivalent to 2.67$\pm$1.37 mas/yr. Assuming spherical expansion, this corresponds to an expansion velocity of V$_{exp} \sim $ 18 km/s in the OH maser shell of W43A, consistent with typical OH/IR expansion velocities \citep{sevenster2002}. Thus, even though the maser morphology indicates a spherical shell is unlikely, there is no strong indication for fast, non-spherical expansion of the OH maser region.

\section{Conclusions}\label{conclusion}

The non-spherical shape of PNe is thought to be related to outflows already generated during the AGB phase. Magnetic fields are considered as collimating agents of the jets around evolved stars. The magnetic field and jet characteristics of W43A, an evolved star in transition to a PN, have been previously reported from H$_{2}$O maser polarization observations. Our GBT observations reveal a magnetic field strength $B_{\rm ||}$ changing sign from $31\pm8$~mG to $-24\pm7$~mG across the H$_2$O masers in the red-shifted lobe of the W43A precessing jet. We observed the OH  maser region of the CSE of this star and measured circular and linear polarization. Due to Faraday rotation, we can not determine the magnetic field configuration in the OH maser shell. However, the measured circular polarization, which is attributed to the Zeeman effect, implies a magnetic field of 100 $\mu$G in the OH  maser region of W43A. Our result is consistent with the predicted magnetic field extrapolated from the blue-shifted H$_{2}$O  maser region of W43A, and further confirms that the magnetic field plays an important role in the transition from a spherical AGB star to  a non-spherical PN.

 We considered the motion and expansion of the OH maser region of W43A. The observed expansion is  2.67$\pm$1.37 mas/yr, which corresponds to an expansion velocity of $\sim$18 km/s. While the significantly offset red- and blue-shifted caps of the OH maser shell indicates it is  a-spherical, the measured OH maser motions do not show any signs of a clear bipolar expansion in the plane of the sky.

\begin{acknowledgements}
This research was supported by the ESTRELA fellowship, the EU Framework 6 Marie Curie Early Stage Training programme under contract number MEST-CT-2005-19669. We acknowledge MERLIN staff for their help in the observation and reducing the data.  We thank Anita Richards for helping us with the initial data processing at Jodrell Bank. WV acknowledges support by the \emph{Deut\-sche
    For\-schungs\-ge\-mein\-schaft} through the Emmy Noether Research
  grant VL 61/3-1. 
\end{acknowledgements}

\bibliographystyle{aa} 
\bibliography{13194ref.bbl} 
\end{document}